# Moiré Excitons Correlated with Superlattice Structure in Twisted WSe$_2$/WSe$_2$ Homobilayers

Trond I. Andersen[1,*], Giovanni Scuri[1,*], Andrey Sushko[1,*], Kristiaan De Greve[1,2,#,*], Jiho Sung[1,2], You Zhou[1,2], Dominik S. Wild[1], Ryan J. Gelly[1], Hoseok Heo[1], Kenji Watanabe[3], Takashi Taniguchi[3], Philip Kim[1,4], Hongkun Park[1,2,†], Mikhail D. Lukin[1,†]

[1]Department of Physics, Harvard University, Cambridge, MA 02138, USA
[2]Department of Chemistry and Chemical Biology, Harvard University, Cambridge, MA 02138, USA
[3]National Institute for Materials Science, 1-1 Namiki, Tsukuba 305-0044, Japan
[4]John A. Paulson School of Engineering and Applied Sciences, Harvard University, Cambridge, MA 02138, USA
*These authors contributed equally to this work.
#Current affiliation: imec, 3001 Leuven, Belgium
†To whom correspondence should be addressed: lukin@physics.harvard.edu, hongkun_park@harvard.edu

**Moiré superlattices in twisted van der Waals materials constitute a promising platform for engineering electronic and optical properties. However, a major obstacle to fully understanding these systems and harnessing their potential is the limited ability to correlate the local moiré structure with optical properties. By using a recently developed scanning electron microscopy technique to image twisted WSe$_2$/WSe$_2$ bilayers, we directly correlate increasing moiré periodicity with the emergence of two distinct exciton species. These can be tuned individually through electrostatic gating, and feature different valley coherence properties. Our observations can be understood as resulting from an array of two intralayer exciton species residing in alternating locations in the superlattice, and illuminate the influence of the moiré potential on lateral exciton motion. They open up new avenues for controlling exciton arrays in twisted TMDs, with applications in quantum optoelectronics and explorations of novel many body systems.**

Engineered moiré superlattices arising from lattice mismatch or relative twist angle between layers can induce periodic potentials for charge carriers and excitons. While conceptually related to quantum simulation experiments *e.g.* in optical lattices (*1*), these systems feature on-chip integrability, electronic tunability, and small (nm) length scales. Moiré superlattices in semiconductor van der Waals (vdW) materials, such as transition metal dichalcogenides (TMDs), are attractive due to the presence of optically active, tightly bound excitons (*2*). Theoretical studies predict that these systems can be used to create topologically nontrivial states (*3, 4*), atomically thin mirrors for quantum optical applications (*5*), and Hubbard model simulators (*6*) analogous to trapped atom arrays (*7*). Recent experiments have explored the effects of moiré patterns in heterobilayers of TMDs, including $MoSe_2/WSe_2$ (*8, 9*), $MoSe_2/WS_2$ (*10*), $MoSe_2/MoS_2$ (*11*), and $WS_2/WSe_2$ (*12*). Photoluminescence and absorption measurements in those experiments suggest that the periodic moiré potential can cause trapping of interlayer exciton states (*8, 9*), hybridization between bands in opposite layers (*10*), and multiple intralayer exciton states (*12*). Most recently, there has been growing interest in twisted TMD homobilayers, where the moiré lattice is governed only by the relative twist angle and not lattice constant mismatch, enabling experimental access to a larger range of superlattice length-scales. This system was recently shown to exhibit superconductivity (*13*), and theoretically predicted to display topological insulator behavior as well as two-orbital Hubbard physics (*4, 14*).

A major experimental challenge in this field involves correlating the optical response with measurements of the spatially varying atomic registry in the moiré superlattice. Although direct imaging of the small (≲100 nm) moiré domains can be achieved through transmission electron microscopy (TEM) (*12*) or scanning probe microscopy (*15*), such imaging techniques require special sample preparation that has previously prevented optoelectronic measurements of the same device. To address this challenge, we employ a recently developed versatile scanning electron microscopy (SEM) technique (*16*) to image the local stacking order of twisted WSe$_2$/WSe$_2$ (t-WSe$_2$/WSe$_2$) homobilayer devices (Fig. 1A, left). The technique is based on measuring secondary electron emission, which depends on the scattering cross section experienced by the incoming primary beam. By placing the device at an angle, different stacking orders exhibit different scattering cross-sections, thus causing contrast between moiré domains. Crucially, this technique, referred to as channeling modulated secondary electron imaging (*16*), is compatible with electrostatically gated devices on regular Si-substrates, thus enabling direct correlation with optoelectronic measurements (Fig. 1A, right).

Figure 1B displays a schematic and optical image of a near-0º t-WSe$_2$/WSe$_2$ device (D1), fully encapsulated in hexagonal BN (hBN) and with a few-layer graphene gate on the bottom. SEM imaging of the device reveals atomic reconstruction of the moiré pattern (Fig. 1C), where the lattices are strained locally to maximize the size of energetically favorable AB and BA (3R) stacking domains (Fig. 1D, right) (*17*). This reconstruction mechanism results in a triangular array of alternating AB/BA regions, as opposed to gradual transitions between the two domain types (Fig. 1D, left) (*17*) – the latter being more prevalent in heterobilayers with intrinsic lattice mismatch and at larger twist angles (*7, 12*).

Critically, the SEM imaging directly demonstrates that the local moiré periodicity varies considerably across and within each of our devices. In particular, an SEM image of a similar hBN-encapsulated and gated device (D2) with a target twist angle of 2.5º shows that the reconstructed moiré wavelength is $\lambda_m$ ~60 nm on the right side of the device, and gradually decreases towards the left (Fig. 2A). Eventually, the moiré wavelength becomes so small (<10 nm) that the reconstructed moiré pattern is no longer observed (left side of the dashed line near the device center).

To explore correlations between the moiré wavelength observed in SEM and the local optical properties, we first collect gate-dependent photoluminescence (PL) spectra from both sides of the device (boxed crosses in Fig. 2A), in locations where $\lambda_m < 10$ nm and $\lambda_m = 35$ nm (Figs. 2B and C, respectively). In both spots, we observe two sets of emission features: the peaks near 1.5 eV are attributed to momentum indirect interlayer excitons based on their energy (*18-20*), whereas the higher energy peaks near 1.7 eV are attributed to K-valley momentum direct intralayer excitons, consistent with previous studies and absorption measurements (SI) (*19-23*). While the interlayer excitons are qualitatively similar in the two spots, the intralayer species exhibit a distinct difference that is most pronounced when the device is p-doped ($V_G < -2$ V). Only one intralayer exciton peak is observed on the left side of the device (Fig. 2B, $\lambda_m < 10$ nm), whereas two peaks split by 37 meV emerge in the right part (Fig. 2C, $\lambda_m = 35$ nm). In what follows, we shall refer to the lower and higher energy peak as peak I and II, respectively. Besides their resonance energies, the doping dependence of peak I and II is also distinct: with increasing hole doping, peak I broadens and intensifies strongly, while peak II is much less affected.

Inspecting more locations (crosses in Fig. 2A) at $V_G = -5$ V, we find that all three points in the left region ($\lambda_m < 10$ nm), exhibit only one dominant intralayer peak, while the two-peak structure persists in all the points in the right part of the device (Fig. 2D). Importantly, moving towards larger moiré domain size ($\lambda_m = 25$ nm → 60 nm), the amplitude of peak I (II) decreases (increases) monotonically.

The observed features in the intralayer exciton emission can be understood by considering the local band structure in the observed AB- and BA-stacked domains of t-WSe$_2$/WSe$_2$. For brevity, we refer to the top layer in BA domains and the bottom layer in AB domains as type I locations (Fig. 3A, maroon squares), and the opposite locations as type II (Fig. 3A, orange squares). In type II locations the W-atoms are directly above or below the Se-atoms in the other layer, while in type I locations they are instead vertically aligned with the hexagon centers (Fig. 3A) (*24*). Due to the different atomic environments, the wavefunctions at the K-point (which are primarily concentrated near the W-atoms) have different energies in these two types of locations.

This gives rise to two effects that split the energies of excitons in type I and II locations. First, the two locations are expected to possess different optical band gaps, due to different valence and conduction band splittings at the K point, $\Delta E_{\text{VB(CB)}} = E^{\text{I}}_{\text{VB(CB)}} - E^{\text{II}}_{\text{VB(CB)}}$ (Fig. 3B). This causes excitons in type II locations to be higher in energy by $\Delta E_0 = \Delta E_{\text{VB}} - \Delta E_{\text{CB}}$. Second, in the doped regimes, the two exciton types are expected to have an additional energy difference because of different interactions (*25*) with the additional charges. In particular, theoretical calculations predict that type I locations have a higher valence band maximum (VBM) at the K-point than type II locations ($\Delta E_{\text{VB}} > 0$, Fig. 3B), and should therefore be preferentially p-doped (Fig. 3A) (*4, 26*). Thus, excitons in hole-doped type I locations can bind strongly to a nearby hole to form charged excitons with reduced energy, while excitons in the neutral type II locations only interact weakly with the holes in the other layer. Based on the stronger hole-doping effects on the type I excitons, we propose that the type I and II excitons reside in type I and II locations, respectively. Furthermore, since it is the lower-energy (type I) exciton that becomes hole-doped, we infer that $\Delta E_{\text{CB}} < \Delta E_{\text{VB}}$ (Fig. 3B), as was also predicted for 3R-stacked bilayer $MoS_2$ (*26*).

The higher energy (type II) peak is only expected to be observed if the excitons are unable to move to a lower energy state in a doped region before recombining ($\tau_X \sim$1-10 ps, Fig. 3C). While interlayer tunneling is strongly suppressed at the K-point in AB/BA-stacked bilayers (*4, 27*), moving laterally to an oppositely stacked domain within the same layer is a faster process (*28, 29*), especially for small domains (Fig. 3D). Assuming a thermalized ensemble of excitons, we estimate the distance that an exciton moves before recombining as $\ell = \sqrt{k_\text{B}T/M_\text{X}} \cdot \tau_\text{X} \sim 10-100$ nm, where $k_B$ is the Boltzmann constant, $M_\text{X} \sim 0.8\,m_\text{e}$ (*30*) is the exciton translational mass, and *T*=4 K is the exciton temperature (SI). The estimated $\ell$ is consistent with the observed decrease (increase) in amplitude of peak I (II) as the domains increase from $\lambda_\text{m} = 25$ nm to $\lambda_\text{m} = 60$ nm across the device (Fig. 2D). We thus interpret the correlations between the PL spectra and $\lambda_\text{m}$ as being due to lateral movement of type II excitons to the potential minima (type I locations), where they relax into lower-energy type I excitons.

Similar doping dependent behavior as in Fig. 2C was observed in several other hBN-encapsulated, near-0° twisted WSe$_2$/WSe$_2$ devices (SI), including D1, which was also confirmed to have a reconstructed moiré pattern by SEM imaging (Fig. 4A, inset). Conducting PL measurements in a spot where $\lambda_\text{m} = 70$ nm, we again observe two peaks with stronger hole-doping effects on the lower-energy exciton, and a very similar intralayer exciton splitting (36 meV) in the far p-doped regime. Device D1 shows more pronounced features in the intrinsic and electron-doped regimes, allowing us to compare the peak splittings in all doping regimes. While peak I red-shifts much more than peak II at the onset of the p-doped regime ($E_\text{h}^\text{I} = 20$ meV, $E_\text{h}^\text{II} \sim 0$ meV), peak II exhibits a larger red-shift in the n-doped regime ($E_\text{e}^\text{II} = 16$ meV, $E_\text{e}^\text{I} = 11$ meV; green arrows in Fig. 4A), thus reducing the splitting to only 10 meV (SI). The observed doping effects in PL (also in absorption; SI) are consistent with recent experimental and theoretical studies suggesting that the electrons are less layer-localized than the holes because the global conduction band minimum is at the Q-point instead of the K-point (*18, 24, 31*). Both exciton species are therefore expected to interact with the electrons. Finally, we obtain a splitting in the neutral regime of $\Delta E_0 = 15 \pm 3$ meV (all measured devices are within this range; SI). The extracted value is smaller than theoretical calculations of the valence band splitting in 3R-stacked bilayer WSe$_2$ ($\Delta E_{VB} = 92$ meV in Ref. (*4*)), suggesting a staggered band alignment at the domain boundaries (Fig. 3B).

In order to substantiate our theoretical interpretation of the two exciton peaks and to further explore their nature, we conduct polarization-resolved photoluminescence spectroscopy. By illuminating with linearly polarized light, we excite a superposition of K and K' valley excitons (excited by $\sigma^+$ and $\sigma^-$ photons, respectively), and subsequently measure the exciton valley coherence by comparing the parallel and perpendicular emission components (*19, 32*). In Figure 4B, we plot the degree of linear polarization, DOLP $= (I_\parallel - I_\perp)/(I_\parallel + I_\perp)$, where $I_{\parallel(\perp)}$ is the intensity of emission with parallel (perpendicular) polarization. The lower-energy (type I) exciton has almost zero DOLP in the p-doped regime, while the higher-energy (type II) exciton exhibits a DOLP of up to 18% (Fig. 4B-C). When entering the neutral regime, the DOLP of the type I exciton increases strongly and exceeds that of type II, which exhibits much less change. In the n-doped regime, the DOLP of both exciton types is found to be even higher, reaching 44% (Fig. 4B and D).

The observed valley coherence properties are consistent with our model of the exciton structures presented in Fig. 3. In the case of type I excitons in the p-doped regime, the additional hole must be in the opposite valley of the exciton due to the Pauli exclusion principle (Fig. 4E). Thus, the hole becomes correlated with the emitted photon, causing decoherence and near-zero DOLP, as observed in our experiment (*33*). For type II excitons, on the other hand, the additional holes are in the opposite layer compared to the exciton and thus carry no information about its valley, allowing for higher coherence than that of type I (Fig. 4F) (*19, 32, 33*). In the intrinsic regime, there are no additional holes, and the DOLP of both exciton types is therefore non-zero. Finally, in the n-doped regime, additional electrons first fill up the Q-valley, while the observed momentum direct intralayer exciton is still expected to be at the K-point. Since the exciton and resident electron are in different valleys, there is no Pauli exclusion principle at play, allowing for non-zero DOLP regardless of which layer the exciton resides in (Fig. 4G). Our model is thus sufficient to explain all observed features qualitatively, however further theoretical studies that include effects of exciton lifetime and additional valley decoherence mechanisms can provide a more quantitative understanding of the DOLP.

Our observations can be understood as resulting from the emergence of spatially alternating exciton species in the reconstructed moiré superlattice, with distinct gate tunability and valley coherence properties. We note that although the bilayer system as a whole is expected to exhibit the same optical response in AB and BA domains, the two layers can be treated independently due to strongly suppressed interlayer tunneling in AB/BA domains at the K point (*4*). Furthermore, a modest vertical electric field (~0.1 V/nm) can lift the degeneracy in AB/BA optical response in the doped regimes. Our experiments therefore indicate that this system can be an attractive solid-state platform for engineering arrays of emitters, and open the door to new quantum optoelectronic applications (*3-5, 7*). In particular, the large range of attainable moiré wavelengths in homobilayers, combined with electrostatic control of the individual regions, can be used to develop new types of moiré-based metasurfaces (*5*). These properties also allow for tuning both exciton-exciton interactions and tunneling between domains, two essential parameters in the Hubbard model (*6, 12*). Specifically, we have shown that the splitting between type I and II excitons can be electrostatically tuned from 36 meV in the p-doped regime to 10 meV in the n-doped regime. One particularly intriguing direction involves realizing one-dimensional localized exciton states, where the electron and hole reside at opposite sides of the sharp moiré domain boundaries in twisted homobilayers. Enabled by the staggered band alignment, such states have been predicted to exhibit large in-plane dipole moments, as well as quantum confinement effects in 10-100 nm sized triangles (*34*). Besides featuring triangular domains with ideal band alignment and length scales, our platform provides regular arrays of such systems, which could permit tunable interactions. This may be utilized for engineering strongly correlated states with strong optical response, potentially enabling the realization of interacting excitons for applications such as quantum nonlinear optics (*5*).

**Acknowledgments:** We thank Bernhard Urbaszek for helpful discussions. **Funding:** We acknowledge support from the DoD Vannevar Bush Faculty Fellowship (N00014-16-1-2825 for H.P., N00014-18-1-2877 for P.K.), NSF (PHY-1506284 for H.P. and M.D.L.), NSF CUA (PHY-1125846 for H.P. and M.D.L.), AFOSR MURI (FA9550-17-1-0002), ARL (W911NF1520067 for H.P. and M.D.L.), the Gordon and Betty Moore Foundation (GBMF4543 for P.K.), ONR MURI (N00014-15-1-2761 for P.K.), and Samsung Electronics (for P.K. and H.P.). All fabrication was performed at the Center for Nanoscale Systems (CNS), a member of the National Nanotechnology Coordinated Infrastructure Network (NNCI), which is supported by the National Science Foundation under NSF award 1541959. K.W. and T.T. acknowledge support from the Elemental Strategy Initiative conducted by the MEXT, Japan and the CREST (JPMJCR15F3), JST. A.S. acknowledges support from the Fannie and John Hertz Foundation and the Paul & Daisy Soros Fellowships for New Americans. **Author contributions:** T.I.A., G.S., A.S., K.D.G., H.P. and M.D.L. conceived the project. T.I.A., G.S., A.S. and K.D.G. designed and performed the experiments, analyzed the data, and wrote the manuscript with extensive input from the other authors. Y.Z. and J.S. assisted with optical measurements. T.I.A.,

G.S., J.S. and Y.Z. fabricated the samples. A.S. and K.D.G performed SEM imaging. T.I.A., G.S., A.S., K.D.G and D.S.W. developed the theoretical model. H.H. grew the TMD crystals. T.T. and K.W. grew the hBN crystals. P.K., H.P., and M.D.L. supervised the project. **Competing interests:** The authors declare no competing interests. **Data and materials availability:** All data needed to evaluate the conclusions in the paper are present in the paper and the supplementary materials.

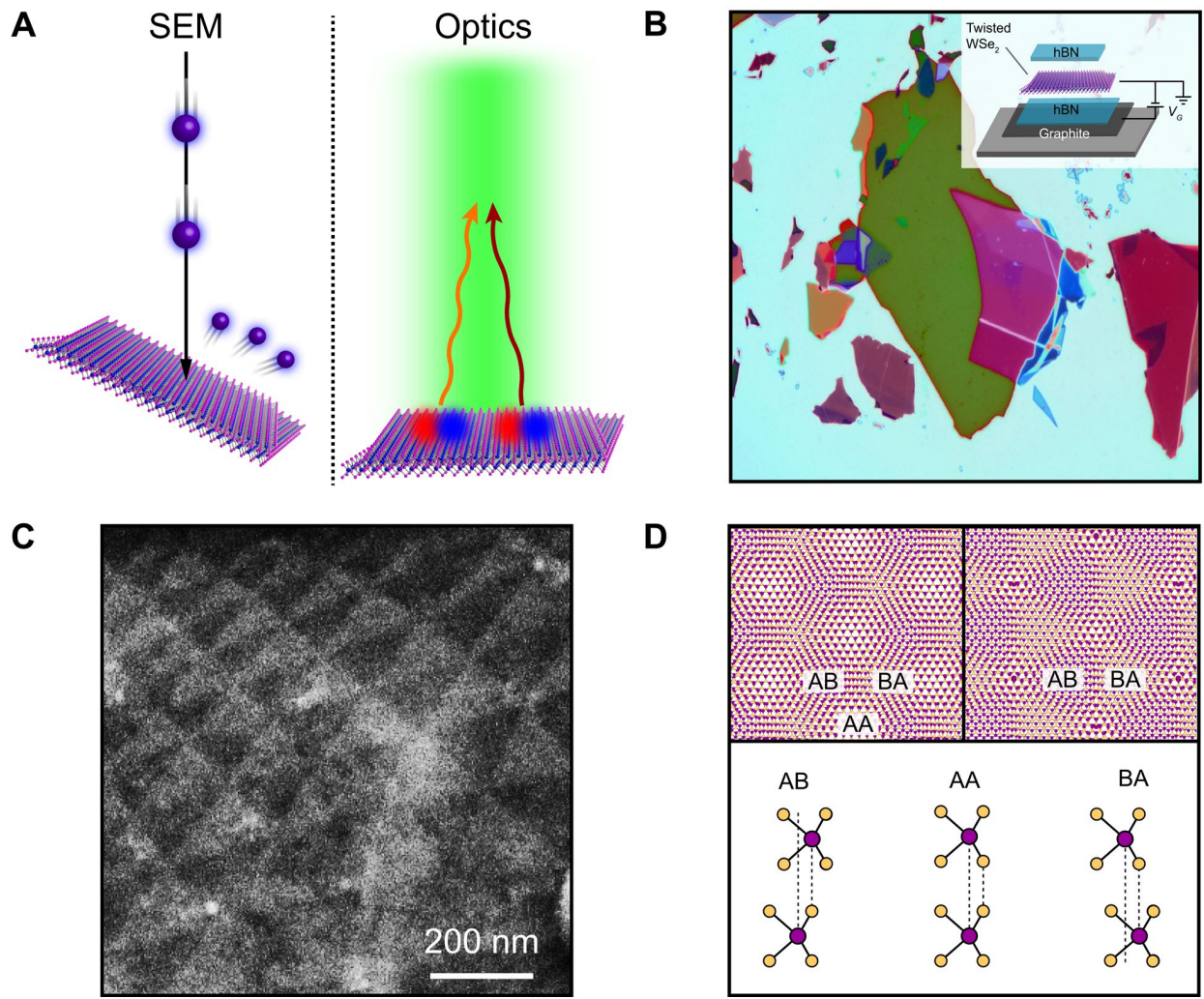

**Fig. 1: Moiré superlattices in twisted homobilayers.** A) Schematic of SEM imaging (left) and optical measurements (right), here correlated in the same device. Large and small electrons indicate primaries and secondaries, respectively. B) Schematic and optical image of hBN-encapsulated twisted WSe$_2$ bilayer (device D1). C) SEM image of device D1, showing a reconstructed moiré pattern with triangular AB and BA stacking domains. D) Top: Top view schematic of moiré pattern with and without reconstruction (right and left). Bottom: Side view of AB, AA and BA stacking orders. Purple: W, yellow: Se.

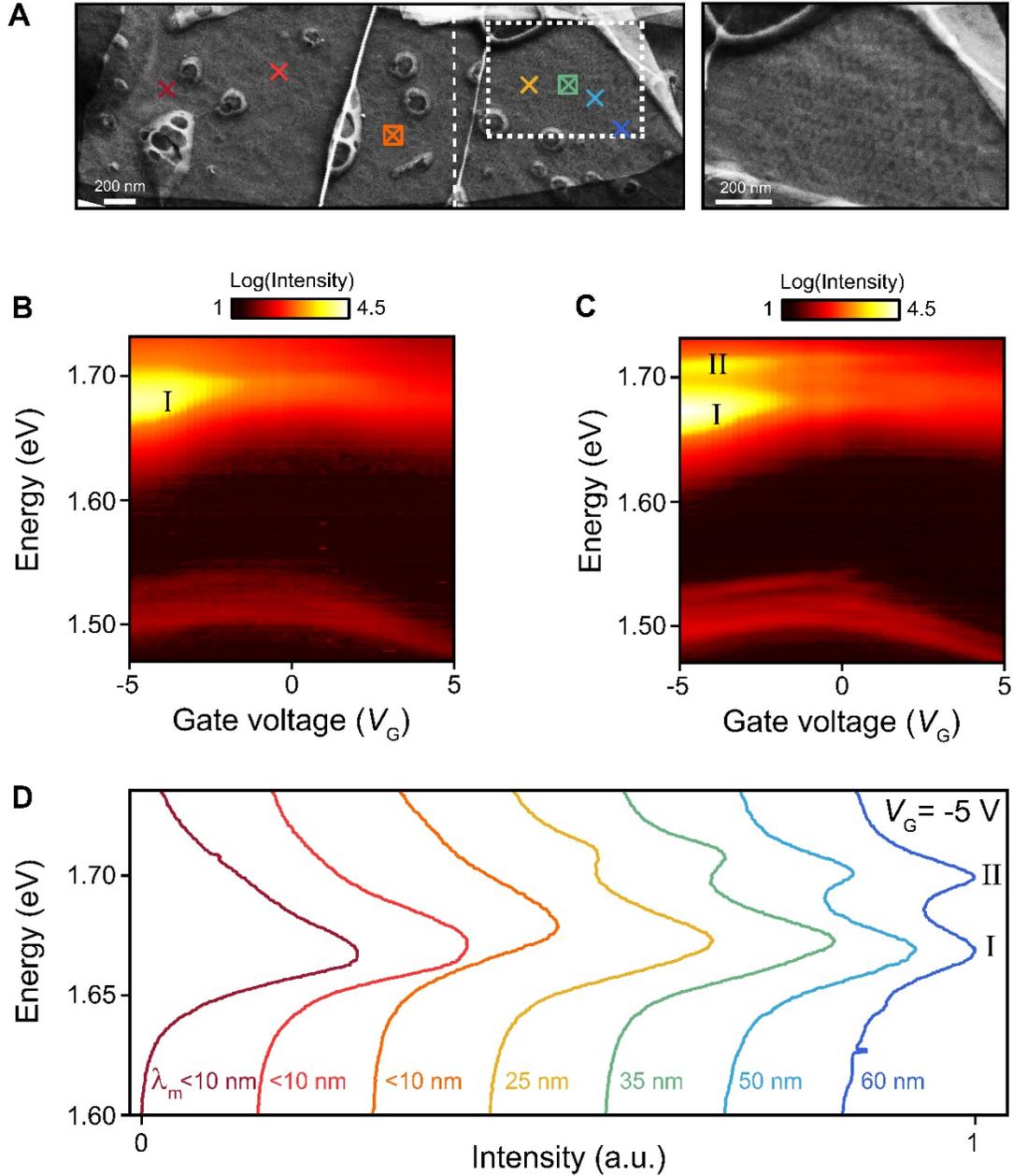

**Fig. 2: Correlating local superlattice structure and optical properties.** A) Left: SEM image of device D2, displaying reconstructed moiré pattern. $\lambda_m$ increases from <10 nm to 60 nm from left to right. The moiré pattern is not visible left of the vertical dashed line. Right: Zoom-in of area indicated by dotted lines. B-C) Gate-dependent PL spectra collected from locations with $\lambda_m < 10$ nm (B) and $\lambda_m = 35$ nm (C), indicated by boxed crosses in (A). While the left region only exhibits one intralayer peak in the p-doped regime, the right (larger $\lambda_m$) region exhibits two peaks. D) Raw PL spectra collected from the locations marked with crosses in (A), at $V_G = -5$ V. The spectra are offset for clarity. The amplitude of peak I(II) decreases (increases) with growing $\lambda_m$.

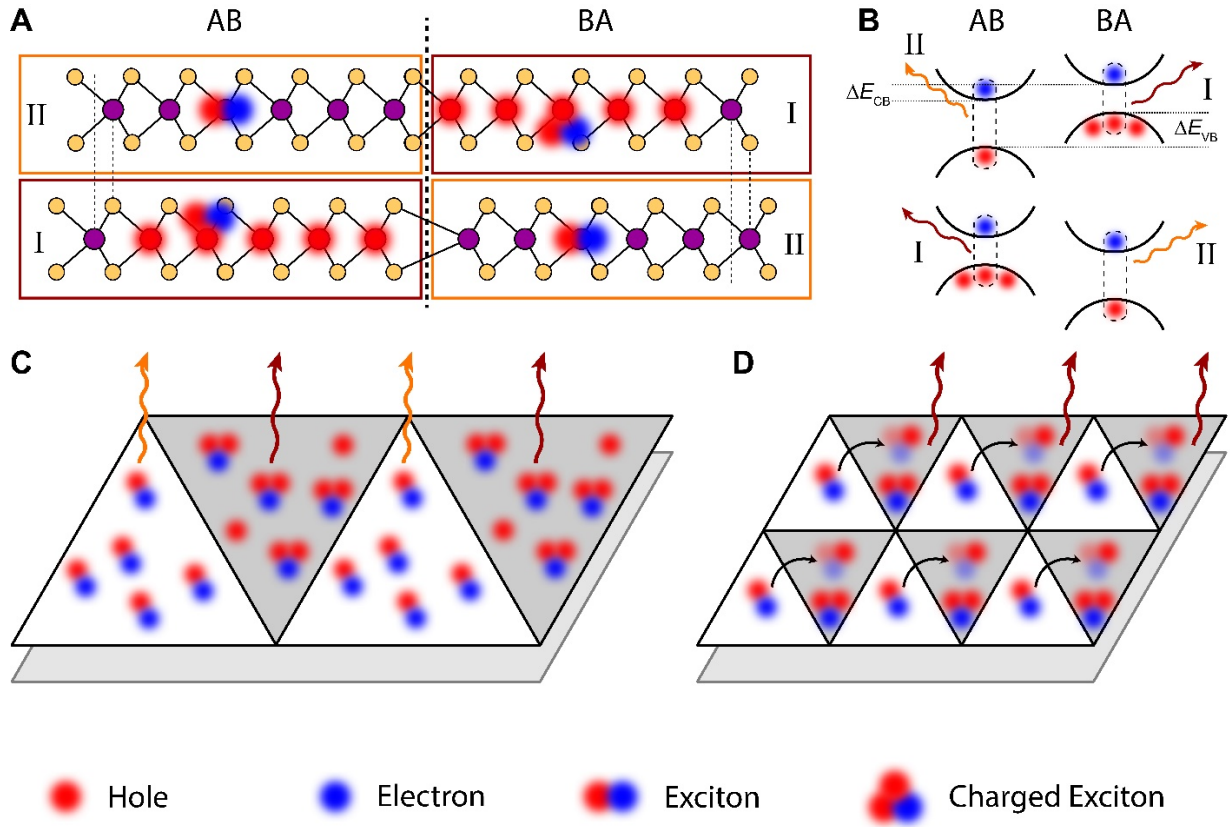

**Fig. 3: Periodic exciton and doping landscape in reconstructed superlattice.** A) Side view of AB and BA domains. Maroon and orange boxes indicate type I and II locations, respectively. In type I locations, excitons have lower energy due to stronger interactions with holes and a smaller optical band gap. B) Schematic of band structure at the K-point in AB and BA domains, for top and bottom layers. The BA (AB) domains have a higher VBM in the top (bottom) layer, and are thus preferentially p-doped. C-D) Schematic of the two-dimensional triangular exciton array. For clarity, only excitons in the top layer are shown. When $\lambda_m$ is large (C), type II excitons are unable to move to other domains before recombining, resulting in an array of spatially alternating emission energies. In the case of smaller $\lambda_m$ (D), type II excitons can move to other domains to form lower-energy (type I) excitons (final positions indicated by faded exciton), causing predominantly type I emission.

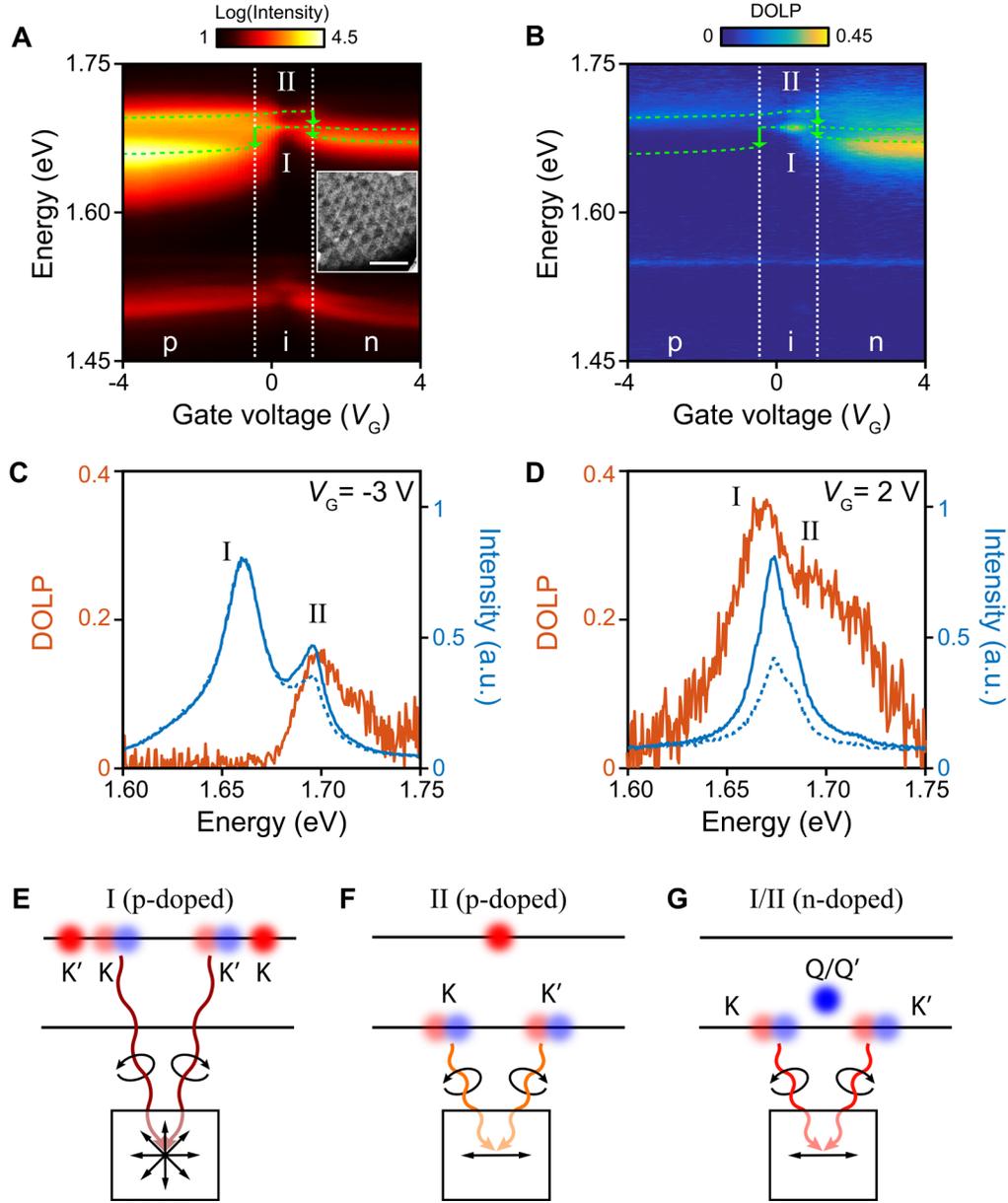

**Fig. 4: Valley coherence of different exciton species in the triangular array.** A) Gate-dependent PL spectra from device D1. Green dashed lines show fitted peak positions (SI). Inset: SEM image of measured spot. Scale bar: 200 nm. B) Gate-dependent DOLP, showing that type I excitons have nearly zero DOLP in the p-doped regime. C-D) Co- and cross-polarized PL (blue solid and dashed curves), and corresponding DOLP (orange) at $V_G = -3$ V (C) and $V_G = 2$ V (D). E-G) Schematics of K/K' exciton recombination. Black horizontal lines indicate top and bottom layer. E) Since the resident hole must be in the opposite valley of the type I exciton, it becomes correlated with the photon and breaks coherence. F) For the type II exciton, the resident hole is in a different layer and carries no information about the exciton valley, allowing for non-zero DOLP. G) DOLP is non-zero for both exciton types in the n-doped regime, because the electron is in a different valley (Q/Q') than the exciton (K/K').

# Supplementary Materials for "Moiré Excitons Correlated with Superlattice Structure in Twisted WSe$_2$/WSe$_2$ Homobilayers"

Trond I. Andersen[1,*], Giovanni Scuri[1,*], Andrey Sushko[1,*], Kristiaan De Greve[1,2,#,*], Jiho Sung[1,2], You Zhou[1,2], Dominik S. Wild[1], Ryan J. Gelly[1], Hoseok Heo[1], Kenji Watanabe[3], Takashi Taniguchi[3], Philip Kim[1,4], Hongkun Park[1,2,†], Mikhail D. Lukin[1,†]

[1]Department of Physics, Harvard University, Cambridge, MA 02138, USA
[2]Department of Chemistry and Chemical Biology, Harvard University, Cambridge, MA 02138, USA
[3]National Institute for Materials Science, 1-1 Namiki, Tsukuba 305-0044, Japan
[4]John A. Paulson School of Engineering and Applied Sciences, Harvard University, Cambridge, MA 02138, USA
*These authors contributed equally to this work.
#Current affiliation: imec, 3001 Leuven, Belgium
†To whom correspondence should be addressed: lukin@physics.harvard.edu, hongkun_park@harvard.edu

**Materials and Methods**

Device fabrication

Flakes of hBN, graphene and WSe$_2$ were first mechanically exfoliated from bulk crystals onto Si wafers (with 285 nm SiO$_2$). Few-layer graphene and monolayer WSe$_2$ were identified optically, while the thickness of hBN flakes was determined with atomic force microscopy. Next, the heterostructures were assembled with the dry-transfer method, using the tear-and-stack technique (*35*) to form twisted bilayer WSe$_2$. Electrical contacts to both the WSe$_2$ and the graphite gates were defined with e-beam lithography and deposited through thermal evaporation (10 nm Cr + 90 nm Au).

Methods

Photoluminescence measurements were conducted in a 4 K cryostat (Montana Instruments), using a custom-built 4f confocal setup with a 0.75 NA objective and a 660 nm Thorlabs diode laser. Two galvo mirrors were used to control the excitation and collection positions on the sample, and electrostatic gating was performed with Keithley multimeters. The DOLP was measured with polarizers in the excitation and collection paths. To eliminate any polarization-dependent system response, all DOLP measurements included excitation with (and collection of) both vertically and horizontally polarized light. The DOLP was then calculated from:

$$\text{DOLP} = \frac{r-1}{r+1}, \ r = \frac{\sqrt{I_{\text{hh}} \cdot I_{\text{vv}}}}{\sqrt{I_{\text{vh}} \cdot I_{\text{hv}}}},$$

where the subscripts indicate the polarization of the excitation and collection, respectively (h: horizontal, v: vertical).

SEM imaging of the reconstructed moiré pattern was done with a Zeiss Field Emission Scanning Electron Microscope Ultra Plus. Detailed methods are presented in Ref. (*16*).

**Supplementary Text**

Extracting energy splitting between the exciton species

In order to extract the different energy scales, we fit the gate-dependent PL spectra with a double Lorentzian of the form:

$$I(E) = \frac{A_\text{I}}{(E-E_\text{I})^2/\Gamma_\text{I}^2+1} + \frac{A_\text{II}}{(E-E_\text{II})^2/\Gamma_\text{II}^2+1},$$

where $A_\text{I(II)}$, $E_\text{I(II)}$ and $\Gamma_\text{I(II)}$ are the amplitude, resonance energy and linewidth of peak I(II). In some devices, the doped and neutral excitons are found to coexist at the transition between the doping regimes. Moreover, in D2 the charged type I exciton peak extends into the intrinsic regime, likely due to localized charge (defect) states, as previously observed (*21, 36*). In these cases, we fit the spectra with a triple Lorentzian.

Fig. S1 shows fitting examples for device D1 and D2 in the three doping regimes, and Fig. S2 displays the full gate dependence of the extracted values of $E_\text{I}$ and $E_\text{II}$. The red-shifts, $E_{\text{h(e)}}^{\text{I(II)}}$, of the type I(II) exciton due to interactions with holes (electrons) are extracted from the change in $E_\text{I}$ and $E_\text{II}$ at the onset of doping (Fig. S2).

Optical measurements of additional devices

Fig. S2 displays gate-dependent photoluminescence spectra (top) and extracted peak splittings (bottom) from the four twisted $WSe_2/WSe_2$ devices studied in our work, which all show very similar behavior. We here list the main common features, which are explained in the main text:

-In the p-doped regime, the type I exciton intensifies, broadens and red-shifts more than the type II exciton.

-At the onset of the n-doped regime, both exciton types red-shift, but the type II exciton red-shifts more.

-The splittings are ($\pm 3$ meV) 35 meV, 15 meV and 10 meV in the p-doped, intrinsic and n-doped regimes, respectively.

Additionally, we show DOLP measurements of device D3 in Fig. S3, displaying the same behavior as D1 presented in the main text.

Lateral exciton motion model

Since the intralayer excitons are short-lived ($\tau_\text{X} \sim 1-10$ ps), the lifetime is expected to be comparable to the mean free scattering time, $\tau_\text{mfp}$. Their motion should therefore be between the diffusive ($\ell \propto \sqrt{\tau_X \tau_{mfp}}$) and ballistic ($\ell \propto \tau_X$) regimes, and the two pictures give similar estimates ($\sqrt{\tau_X \tau_{mfp}} \sim \tau_X$).

We also note that the exciton temperature can be slightly elevated due to laser heating. Nevertheless, since we use a low-power continuous wave laser and have not observed any heating-induced exciton peak shifts, we expect that the exciton temperature remains below 10 K, thus not impacting the conclusions in the main text.

Comparison of PL and absorption spectroscopy

We also conduct gate-dependent absorption spectroscopy of device D4, shown as the derivative of differential reflection in Fig. S4A (PL spectra are shown in Fig. S4B for comparison). While interlayer exciton absorption is too weak to be observed, we observe strong absorption features from the intralayer excitons (*22, 25*). This is consistent with our assignment of the intralayer excitons to the momentum direct K-K transition.

The doping dependence observed in absorption measurements is also consistent with PL. While the type I exciton exhibits much stronger doping effects (blue-shift and weakening of neutral state) than the type II exciton at the onset of the p-doped regime, the type II exciton exhibits stronger doping effects at the onset of the n-doped regime. In the n-doped regime, the type I exciton also shows clear signs of doping, as expected from the fact that the electrons are more delocalized than the holes. The absorption spectra also show the red-shifted charged exciton states that were observed in PL.

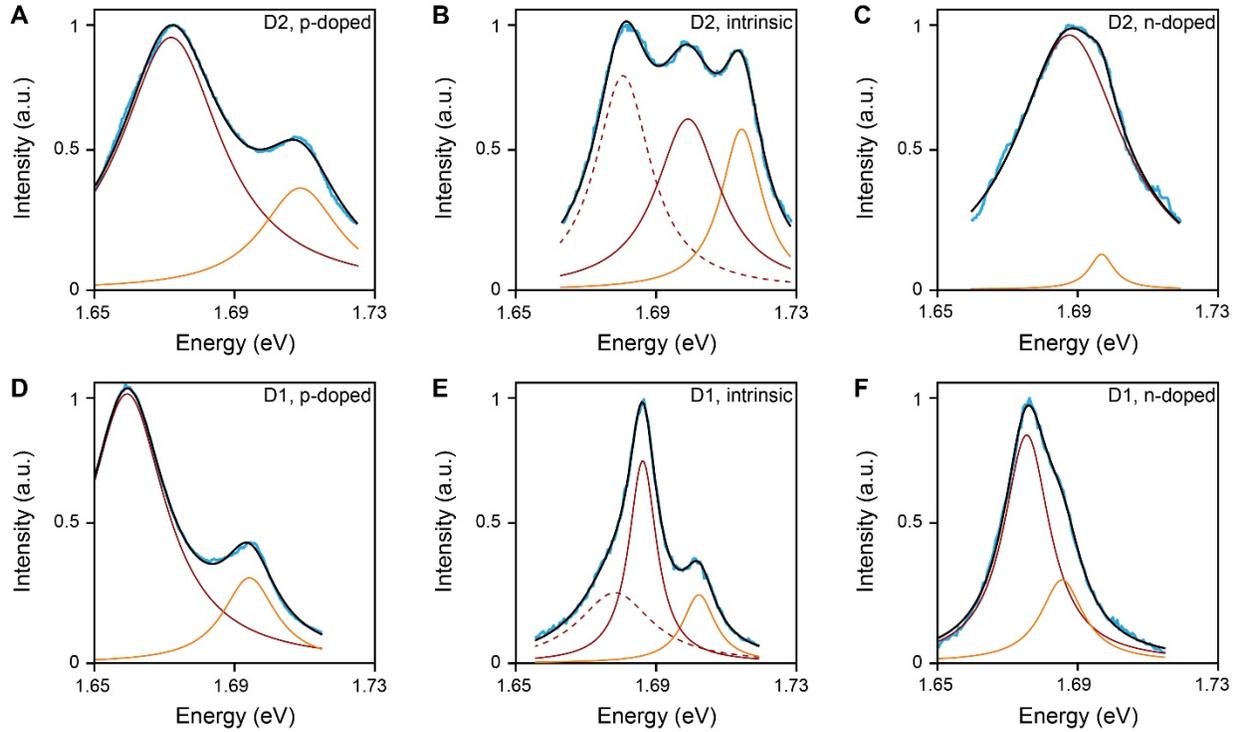

**Fig. S1: Peak fitting in different doping regimes.** A-C) Photoluminescence spectra from device D2 in the (A) p-doped, (B) intrinsic and (C) n-doped regimes (light blue lines). D-F) Same as (A-C), for device D1. In (A), (C), (D) and (F), the spectra are fit with double Lorentzians (black), with contributions from the type I (II) excitons shown in maroon (orange). In (B) and (E), the charged type I exciton is observed even in the neutral regime (dashed maroon), likely due to localized charge defects. We therefore fit with a triple Lorentzian (black). The extracted peak splittings are 37 meV, 15 meV and 10 meV in (A-C), and 36 meV, 16 meV and 10 meV in (D-F).

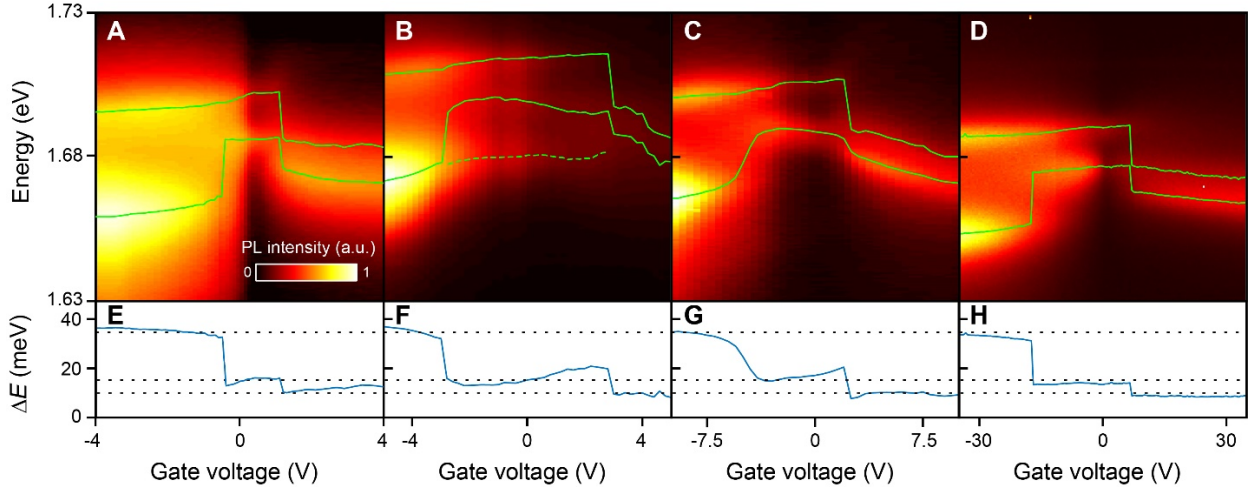

**Fig. S2: Optical measurements of additional devices.** A-D) Gate dependent photoluminescence from devices D1, D2, D3 and D4 (left to right). Green lines show the extracted peak positions from double or triple Lorentzian fits. The dashed green line in (B) indicates that the charged type I exciton emission persists into the intrinsic regime, likely due to localized charge defects. The width of the intrinsic regime varies due to differences in contact quality and hBN thickness. E-H) Peak splitting calculated from green lines in (A-D). Dashed horizontal lines highlight that the splitting is very similar across all four devices.

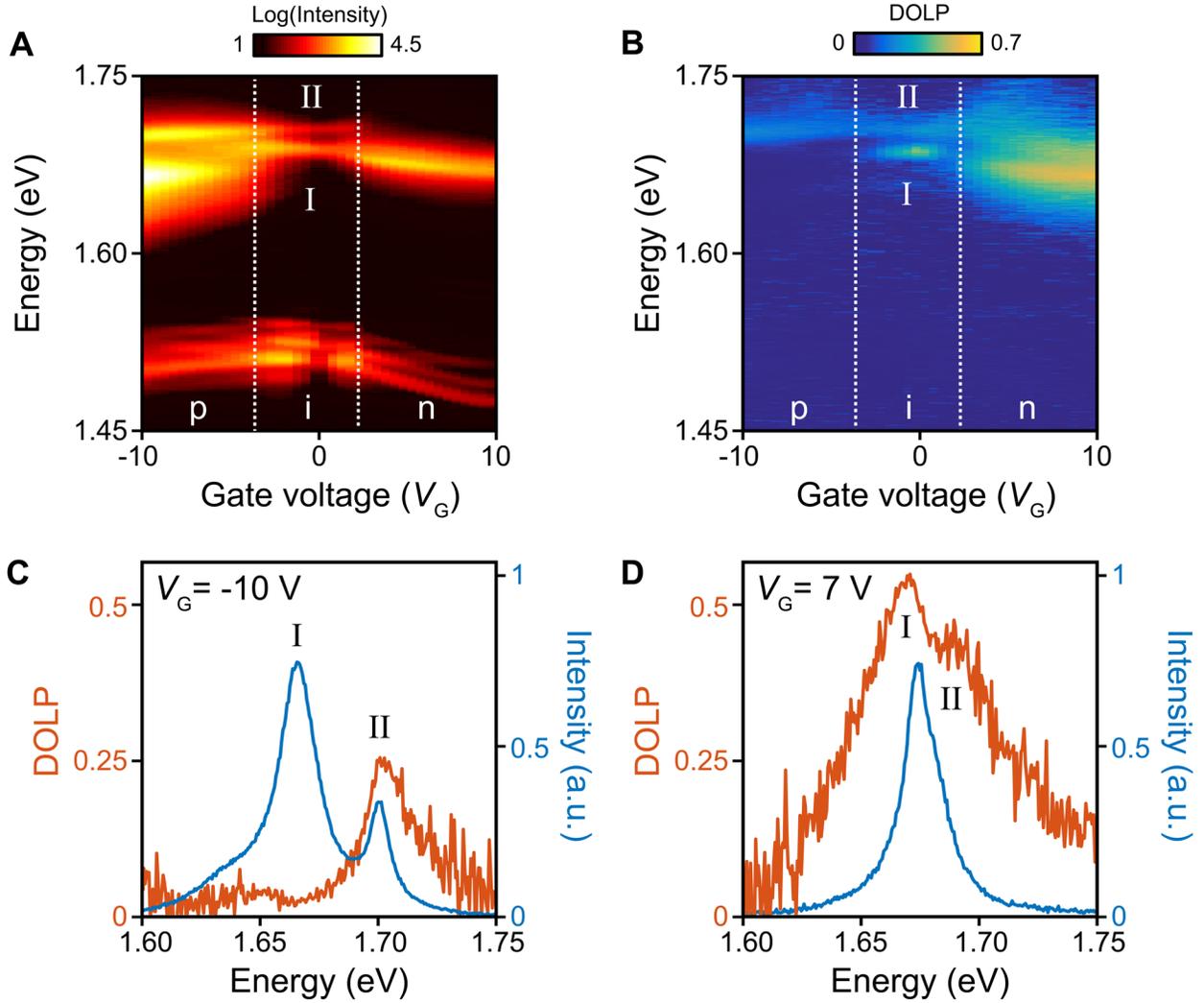

**Fig. S3: DOLP measurements in device D3.** A) Gate-dependent PL spectra from device D3. B) Corresponding DOLP, showing the same behavior as device D1 presented in the main text. C-D) Co- and cross-polarized PL (blue solid and dashed curves), and corresponding DOLP (orange) at $V_G = -10$ V (C) and $V_G = 7$ V (D).

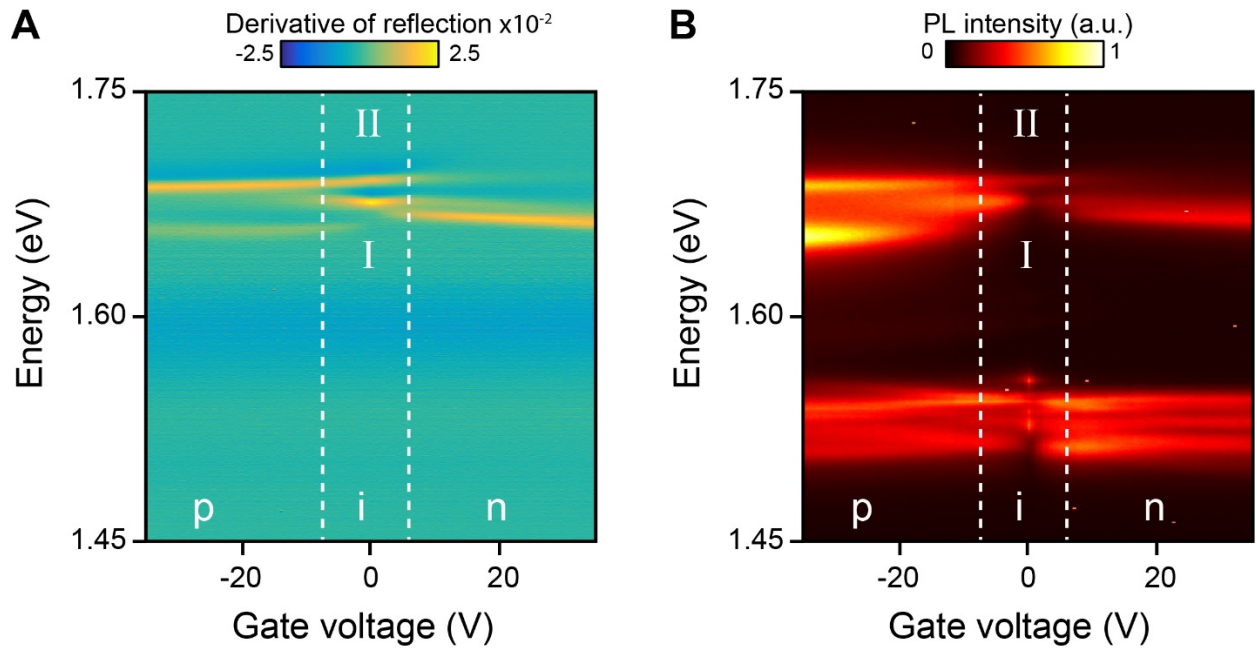

**Fig. S4: Comparison of absorption and photoluminescence spectra.** A) Derivative of differential reflectance vs. gate voltage in device D4, showing strong intralayer exciton absorption, consistent with the momentum direct K-K transition. B) Corresponding gate-dependent photoluminescence spectra from device D4.